\documentstyle[twocolumn,prb,aps,psfig]{revtex}

\newcommand{\rK}[1]{\left( #1 \right)}           
\newcommand{\eK}[1]{\left[ #1 \right]}           
\newcommand{\gK}[1]{\left\{ #1 \right\} }        
\newcommand{\dddot}[1]{\overdots{#1}}

\newcommand{\pdiff}[2]{\frac{\partial #1}{\partial #2}}

\newcommand{\om}{\omega}
\newcommand{\al}{\alpha}
\newcommand{\vt}{\vartheta}
\newcommand{\de}{\delta}
\newcommand{\la}{\lambda}
\newcommand{\si}{\sigma}
\newcommand{\ka}{\kappa}

\begin{document}
\draft

\title{Magnon modes and magnon-vortex scattering in two-dimensional easy-plane 
   ferromagnets}
\author{B.~A.~Ivanov, \cite{BAI_address}  H.~J.~Schnitzer, and F.~G.~Mertens} 
\address{Physikalisches Institut, Universit\"at Bayreuth, D-95440 Bayreuth, Germany}
\author{G.~M.~Wysin}
\address{Department of Physics, Kansas State University
   Cardwell Hall, Manhattan, Kansas 66506, USA}

\maketitle

\begin{abstract}
We calculate the magnon modes in the presence of a vortex on a circular 
system, combining analytical calculations in the continuum 
limit with a numerical diagonalization of 
the discrete system. The magnon modes are 
expressed by the 
$S$-matrix for 
magnon-vortex 
scattering, as a 
function of the 
parameters and the size 
of the system and for 
different boundary 
conditions. Certain 
quasi-local translational modes are identified with the frequencies which 
appear in the trajectory $\vec{X}(t)$ of the vortex center in 
recent Molecular 
Dynamics simulations of 
the full many-spin 
model. Using these quasi-local modes we calculate the two parameters of a 
$3^{rd}$-order equation of motion for $\vec{X}(t)$. This equation was 
recently derived by a collective variable theory and describes very well the 
trajectories observed in the simulations. Both parameters, the vortex mass 
and the factor in front of $\dddot{\vec{X}}$, depend strongly on the 
boundary conditions.
\end{abstract}
\pacs{75.10.Hk,75.40.Gb,75.40.Mg,02.60.Cb}

\section{Introduction}
Two-dimensional (2D) magnets have been intensively investigated due to 
different reasons: There are several classes of quasi-2D materials for which 
the magnetic interactions within planes of magnetic ions are typically three 
to six orders of magnitude larger than the interactions between the 
planes. These classes include layered magnets (e.g. Rb$_2$CrCl$_4$), 
graphite intercalated compounds (e.g.\ CoCl$_2$), magnetic lipid 
layers (e.g.\ manganese stearate), and high-$T_c$ superconductors. For 
theoreticians 2D magnets with $XY$- or easy-plane symmetry are particularly 
interesting due to the existence of vortices, which have nontrivial 
properties from the viewpoint of certain homotopical groups, see reviews 
\onlinecite{Kosevich90,Baryakhtar93,Ivanov95}.

There are two types of vortices: 
In-plane (IP) ones with all spins lying in the easy plane, and
out-of-plane (OP) ones having non-zero spin components orthogonal to
the easy plane. Both types have a $\pi_1$-topological charge or vorticity
$q=\pm1,\pm2,\ldots$, which determines the directions of the spins in
the easy plane far away from the vortex center.
The OP-vortices have an additional $\pi_2$-topological charge
(Pontryagin invariant, see Ref.\ \onlinecite{Ivanov95}) $Q=-\frac{1}{2}qp$ where
$p$ is integer. $p$ is denoted polarization because its sign determines
to which side of the easy plane the spins point in the vortex center.
Therefore the IP-vortices can be considered as having $p=0$. For
$Q\ne0$ there is a gyrocoupling force, or gyroforce,
which is formally equivalent to the Lorentz force. \cite{Malozemoff79}

Besides the vortices, which are strongly nonlinear excitations, 
there are also magnons. The dynamic properties of
2D easy-plane magnets can be described by a phenomenological
theory  which assumes two ideal gases for the vortices and magnons.
\cite{Mertens87} The vortex-magnon interaction naturally is the next 
question which arises. In particular, one would like to know how the 
dynamics of the vortices is affected by this interaction.

For 1D magnets the soliton-magnon interaction is nearly reflectionless. 
In this case the main effect of the soliton-magnon interaction is the change 
of the magnon density of states, which affects strongly the soliton density in 
thermal equilibrium. \cite{Mikeska91}
For 2D magnets the vortex density can be obtained from the
correlation length by a 
renormalization group approach, \cite{Kosterlitz73,Kosterlitz74,Berezinskii70}
therefore the vortex-magnon interaction is 
not so important for the density. In principle, it can be important for the 
damping force which acts on a vortex (or soliton) in a near-equilibrium 
magnon gas. \cite{Baryakhtar93}  On the other hand, the damping force 
for the 2D case can be obtained by general hydrodynamic theories for the 
relaxation processes in magnets. \cite{Galkina96}

The most important effect of the vortex-magnon interaction seems to be
that for a finite system certain magnon modes are excited due to 
the vortex motion, and vice versa.
Such modes were obtained in recent papers \cite{Wysin94,Wysin95,Ivanov96} 
by a numerical diagonalization for relatively small, circular, discrete 
systems. A calculation in the continuum limit was presented  
for the antiferromagnetic case. \cite{Ivanov96} Analytical investigations 
were done  for
planar vortices in antiferromagnets \cite{Costa92} and ferromagnets. \cite{Pereira96}
These articles demonstrate nontrivial properties of the eigenmodes,
e.g., the presence of quasi-local \cite{Wysin94,Wysin95} or truly local 
\cite{Ivanov96} modes. Moreover, the relevance of these modes for the 
vortex dynamics was shown, in particular an IP-OP transition was predicted 
\cite{Wysin94,Wysin95} and an effective vortex mass was defined and 
calculated using the above modes. \cite{Wysin96} However, for OP-vortices 
all this was 
based on the numerical diagonalization of small systems with fixed 
 (Dirichlet) boundary conditions (b.c.). 

This article deals with a more general theory of vortex-magnon coupling for 
arbitrarily large systems with circular symmetry and general boundary
conditions.
The method consists of a natural combination of analytical calculations
in the continuum limit with numerical diagonalization of 
discrete systems.
The main issue is the calculation of the scattering matrix
for vortex-magnon collisions. \cite{fn1} 
Using these data general formulas for the eigenfrequencies are obtained, as 
a function of the parameters and size of the system, and for different 
boundary conditions.

An important point is a link with a recent collective variable theory 
\cite{Mertens97} in which an equation of motion was derived which 
describes very well the vortex trajectories observed in computer
simulations. This equation is $3^{rd}$ order in time, and is a 
generalization of the Thiele Eq. \cite{Thiele73,Huber82} which is first 
order in time. Using the scattering data the parameters of the 
$3^{rd}$-order equation of motion can be calculated, and agree very well 
with those obtained from simulations. \cite{Mertens97}

\section{The model and elementary excitations}
We consider the classical 2D model of a Heisenberg ferromagnet
(FM) with the Hamiltonian
\begin{equation}
\label{model}
   \hat H = -J \sum_{(\vec n,\vec n')}\eK{ \vec S_{\vec n}\vec S_{\vec n'}
       - (1-\lambda) S_{\vec n}^z S_{\vec n'}^z }.
\end{equation}
Here $J>0$ is the exchange integral, and $0 \leq \lambda<1$ 
describes easy-plane anisotropy. The spins $\vec S$ are classical
vectors on a square lattice with the lattice constant $a_0$.
$\vec n,\vec n'$ denote nearest-neighbor lattice sites.
Our main interest lies in the small anisotropy case which
corresponds to $1-\lambda\ll 1$.

A continuum model for FMs can be derived from (\ref{model})
in the usual way defining the unit vector of magnetization
 as a function of continuous variables $\vec r$ and $t$,
i.e.\ $\vec m(\vec r,t) = \vec S_{\vec n}(t)/S$.
The dynamical equations for $\vec m$ have the form of the
well-known Landau-Lifshitz equation, see 
Refs.\ \onlinecite{Kosevich90,Baryakhtar93}. 
In usual angular variables 
$[m_x+im_y = \sin\theta\exp(i\phi),m_z = \cos\theta]$ 
they can be written as
\begin{eqnarray}
\label{eq2}
   \nabla^2\theta - \sin\theta\cos\theta
   \eK{(\nabla\phi)^2 - \frac{1}{r_v^2}} &=&
   +\frac{\sin\theta}{cr_v} \pdiff{\phi}{t} \\
\label{eq3}   
\nabla(\sin^2\theta\nabla\phi) &=& -\frac{\sin\theta}{cr_v}
   \pdiff{\theta}{t}
\end{eqnarray}
where $r_v$ and $c$ are the characteristic length scale
and magnon speed, respectively. For the Hamiltonian
(\ref{model}), we have
\begin{equation}
\label{eq4}
   r_v = \frac{a_0}{2}\sqrt{\frac{\lambda}{1-\lambda}}\quad,\quad
   c = 2JSa_0\sqrt{1-\lambda}
\end{equation}
where we set $\hbar = 1$. Note that the Eqs.\ (\ref{eq2},\ref{eq3}) arise
in the long-wave approximation ($a_0|\nabla\vec m| \ll 1$) not only for the 
model we are considering here, but for a whole set of discrete
models, for example, FMs on different kinds of lattices and
FMs with additional single-ion anisotropy.
Merely the expressions for $c$ and $r_v$  change.

For the homogeneous ground state (all spins are parallel and
confined to the easy plane) the 2D model has well-known
magnon excitations with the gapless dispersion law
\begin{equation}
\label{eq5}
   \omega = ck(1+k^2r_v^2)^{1/2},
\end{equation}
where $k=|\vec k|$ and $\vec k$ is the magnon wavevector.

Other well-known excitations are the
OP-vortices, described by the formulas
\begin{equation}
\label{eq6}
\theta = \theta_0(r) \quad , \quad \phi = q\chi + \varphi_0 ,
\end{equation}
where $r$ and $\chi$ are polar coordinates and $\varphi_0$ is an arbitrary
constant. 
The function $\theta_0$ is the solution of a nonlinear ordinary
differential equation which cannot be solved analytically.
\cite{Kosevich90,Baryakhtar93} Numerical integration, however, shows a 
surprisingly good agreement with the data obtained from the numerical 
analysis of the full discrete model (\ref{model}), even for $r_v=1.5a_0$ 
($\lambda=0.9$), Ref.\ \onlinecite{Ivanov96}. This means that at least concerning
static properties strong inequalities like $a_0|\nabla \vec m|\ll1$ can
be replaced by usual ones. As will be seen, the same is correct for some 
dynamical properties, too. $\theta_0$ fulfills the boundary conditions 
$\cos\theta_0\to p$ for $r\to0$ and $\theta_0\to\pi/2$ for 
$r\to\infty$. $q$ and $p$ are topological charges of
vorticity and polarization. In the following OP-vortices with $q=1$
 will be considered; for definiteness we set $p = 1$.

\section{Normal modes on the vortex:  continuum approach}
\label{sec:normal}
We consider small deviations from the static vortex solutions, i.e.
\begin{equation}
\label{eq7}
   \theta = \theta_0(r) + \vt \quad,\quad
   \phi = q\chi + (\sin\theta_0)^{-1} \mu.
\end{equation}
Substituting this in the equations (\ref{eq2}) and (\ref{eq3}) and 
linearizing in $\vt$ and $\mu$ gives the following set of
coupled partial differential equations:
\begin{eqnarray}
\label{eq8}
   \eK{-\nabla_x^2+V_1(x)}\vt + \frac{2q\cos\theta_0}{x^2} 
    \pdiff{\mu}{\chi} &=& -\frac{r_v}{c}\pdiff{\mu}{t} \\
\label{eq9}   
\eK{-\nabla_x^2+V_2(x)}\mu - \frac{2q\cos\theta_0}{x^2} 
    \pdiff{\vt}{\chi} &=& +\frac{r_v}{c}\pdiff{\vt}{t},
\end{eqnarray}
where $x = r/r_v$ and $\nabla_x = r_v\nabla$. 
The additional factor $(\sin\theta_0)^{-1}$ in (\ref{eq7}) was introduced
for convenience, because it leads to equations that are symmetric
in $\vt$ and $\mu$ with Schr\"odinger-type differential operators
in front.  The ``potentials'' $V_1(x)$, $V_2$(x)
have the same form as for the antiferromagnet, \cite{Ivanov96}
\begin{equation}
\label{eq10}
   V_1 = \left[\frac{q^2}{x^2}-1\right]\cos 2\theta_0 ,\quad
   V_2 = \left[\frac{q^2}{x^2}-1\right]\cos^2\theta_0 
       - \left[\frac{d\theta_0}{dx}\right]^2,
\end{equation}
but the dynamical parts differ strongly. 

In order to solve (\ref{eq8}, \ref{eq9}) the following ansatz for $\vt$ and 
$\mu$ is appropriate:
\begin{eqnarray}
\label{eq11}
   \vt &=& \sum_n\sum_{m=-\infty}^{+\infty}
      \gK{[f_\al' + i f_\al'']\exp(im\chi + i\om_\al t) + c.c.} \\
\label{eq12}  
 \mu &=& \sum_n\sum_{m=-\infty}^{+\infty}
      \gK{[g_\al' + i g_\al'']\exp(im\chi + i\om_\al t) + c.c.} 
\end{eqnarray}
$\al=(n,m)$ is a full set of numbers labeling the magnon 
eigenstates. 
Substituting this ansatz gives two uncoupled sets of
two equations for the pairs of functions $(f',g'')$ on the one
and $(f'',g')$ on the other side.
However, the equations for $(f'',g')$ can be obtained
from the corresponding equations for $(f',g'')$ simply by replacing
$m\to-m$ and $\om\to-\om$.
Thus, among the four functions $f'$, $f''$, $g'$ and $g''$ only
two are independent. Therefore, instead of (\ref{eq11}, \ref{eq12}) we can use
the simplified ansatz
\begin{eqnarray}
\label{eq13}
   \vt = \sum_n\sum_{m=-\infty}^{+\infty}
      f_{\alpha}(r) \cos(m\chi+\om_{\alpha} t + \de_m) \\
\label{eq14} 
  \mu    = \sum_n\sum_{m=-\infty}^{+\infty}
      g_{\alpha}(r) \sin(m\chi+\om_{\alpha} t + \de_m) 
\end{eqnarray}
with arbitrary phases $\de_m$.
For the functions $f$ and $g$ (the index $\alpha$ is omitted in the following)
we then finally obtain the following two differential equations: 
\begin{eqnarray}
\label{eq15}
   \eK{\frac{d^2}{dx^2} + \frac{1}{x}\frac{d}{dx} - \frac{m^2}{x^2} - V_1}
      f = \eK{\frac{\om r_v}{c} + \frac{2q m\cos\theta_0}{x^2}} g \\
\label{eq16}   
\eK{\frac{d^2}{dx^2} + \frac{1}{x} \frac{d}{dx} - \frac{m^2}{x^2} - V_2}
      g = \eK{\frac{\om r_v}{c} + \frac{2q m\cos\theta_0}{x^2}} f.
\end{eqnarray}
$f$ and $g$ cannot be determined analytically from Eqs.\ (\ref{eq15}, 
\ref{eq16}). The ``potentials'' $V_1(x)$ and $V_2(x)$ in these equations are not
small, and the use of the Born approximation looks inadequate; 
for IP-vortices and $m = 0, \pm 1$ this was already mentioned in
Ref.\ \onlinecite{Pereira96}.
 Only the asymptotic behavior can be calculated. For
$r\to0$ we obtain $g,f \sim r^{|q+m|}$, which describes the
presence of a ``hole'' in the functions $\vt$, $\mu$
at the vortex core for large values of $m$ (see Section 
\ref{sec:results}). Only the case $|m| = 1$ can be considered in a long-wave
approximation, see below (\ref{f=d}-\ref{grleft}).

For the IP-vortex, where $\cos\theta_0$ is zero, the modes corresponding to
$\pm|m|$ are degenerate. As a change of sign in the number
$m$ can also be interpretated as a change of the sense of rotation
of the eigenmode (change of sign in the eigenfrequency $\om$),
physically we have the situation of two independent oscillators
rotating clockwise and counterclockwise with the same frequency (which
can also be combined to give two linear oscillators in independent 
directions). 

For the OP-vortex, however, with $\cos\theta_0\ne 0$, this degeneracy is 
removed due to the presence of the term with $2q m\cos\theta_0/r^2$.
Therefore the two senses of rotations corresponding to positive
and negative frequencies are not equivalent anymore.
The presence of this term is also responsible for the fact that
the Eqs.\ (\ref{eq15}) and (\ref{eq16}) are only invariant under the combined
conjugation of both topological charges 
$(q,p)\to(-q,-p)$. The product $qp$ is proportional to the
magnitude of the so-called gyrovector,
which acts on the vortex like a (self-induced) magnetic field
in $z$-direction for a charged particle. Therefore, physically,
for $|m| \ge 1$ the removal of the $\pm|m|$-degeneracy in the OP-case can
be understood as the effect of a gyroscopic force.

In order to describe the magnon scattering due to the vortex,
we note that without the vortex the eigenvalue problem (EVP) (\ref{eq8}, 
\ref{eq9}) can be reduced to a usual Schr\"odinger EVP with the general 
solution  
\begin{eqnarray}
\label{eq17}
   \mu &=& \sum_m C_m J_m(kr) \sin(m\chi + \om t + \de_m) \\
  \label{eq18}
 \vt &=& \sum_m C_m \frac{k r_v}{\sqrt{1+k^2r_v^2}} J_m(kr) 
                  \cos(m\chi + \om t + \de_m).
\end{eqnarray}
The $J_m$ are Bessel functions, $k$ and $\om$ are connected by the 
dispersion law (\ref{eq5}), and $C_m$ and $\de_m$ are arbitrary constants. 
For a finite circular system with radius $L$ and general boundary conditions 
\begin{equation}
\label{gen_bc}
   \rK{a\mu + b r_v\pdiff{\mu}{r}}_{r=L} = 0\quad,\quad
   \rK{a\vt + b r_v\pdiff{\vt}{r}}_{r=L} = 0
\end{equation}
with arbitrary constants $a$ and $b$, 
the values of the frequencies $\om$ are determined by the roots
of linear combinations of Bessel functions and its derivatives. Note that 
{\it all}
frequencies for the system without a vortex have a $1/L$-dependence 
with respect to the system size. 
For free boundary conditions, i.e.\ $a=0$ in (\ref{gen_bc}), there
is a Goldstone mode (GM) with $\om=0$ due to the rotational symmetry
in spin space of the model. For $a\ne0$ this symmetry is obviously broken. 

In the presence of a vortex, one can use the approximate formula
$f\approx k r_v g/(1+k^2 r_v^2)^{1/2}$ for large distances 
$r\gg r_v$ and then the EVP (\ref{eq15}, \ref{eq16}) reduces to the form
of a SEVP, too. The solutions have the same form as (\ref{eq17}, 
\ref{eq18}), one only has to replace
\begin{equation}
\label{eq20}
   J_m(kr) \to J_m(kr) + \si_m(k) Y_m(kr).
\end{equation}
Here the $Y_m$ are Neumann functions. The quantity $\si_m(k)$ determines 
the intensity of magnon scattering due to the presence of the vortex.
In usual scattering theory for 2D Schr\"odinger equations the 
$S$-Matrix $S_m(k)$ can be represented as 
\begin{equation}
\label{eq21}
   S_m(k) = \frac{1-i\si_m(k)}{1+i\si_m(k)}.
\end{equation}
The values of $\si_m(k)$ are determined by the shape of the solution near
the vortex core, practically at $r/r_v<3\ldots4$.
Due to the structure of the continuum equations (\ref{eq15}, \ref{eq16}), 
$\si_m(k)$ must be
some universal function of $\ka=kr_v$, independent of the concrete
values of the magnetic coupling constants, the system size or
the applied boundary conditions.

$\si_m(k)$ could be calculated directly by solving the set of equations
(\ref{eq15}, \ref{eq16}) using a shooting procedure as described in 
Ref.\ \onlinecite{Ivanov96} for the case of the antiferromagnet. 
The shooting parameter 
is the free constant which appears in the general solution of the set
of two coupled Schr\"odinger-like EVP. In this article we used another
approach and extracted the scattering data from the 
eigenfrequencies $\om_i$ we found numerically for the discrete
2D system in Section \ref{sec:results}. 

In  the next section the general properties of $\si_m(k)$ will be
established for different $m$, and characteristic features of the normal
modes will be discussed. Here we consider only the case $|m| = 1$, for
which the S-matrix in the long-wave approximation can be calculated
analytically. 

This can be done by using the fact that for $k = 0\, (\omega = 0)$ 
and $L \rightarrow \infty$ a non-trivial solution of (\ref{eq15}, \ref{eq16}) 
is known, and by applying a special perturbation theory. This solution
reads $f_0 = d \theta_0/dx, g_0 = - \sin \theta_0/x$, and corresponds
to a vortex displacement (translational Goldstone mode).

In order to construct the asymptotics of such a solution for a small
but finite frequency, we make the ansatz 
\begin{equation}
\label{f=d}
   f = (d\theta_0/dx) [1 + \alpha(x)]\,,\, g= -(\sin\theta_0/x)[1 + 
   \beta(x)],
\end{equation}
where $\alpha(x), \beta(x)$ must be proportional to $\omega$.

Inserting this ansatz into the set of Eqs. (\ref{eq15}, \ref{eq16}), 
multiplying (\ref{eq16}) with $g_0$ and (\ref{eq15}) with $f_0$, and
adding the results, one obtains the equation 
\begin{equation}
\label{left}
   \left\{\left[\beta' \left(\frac{\sin\theta_0}{x}\right)^2 + \alpha' 
   (\theta'_0)^2\right]x\right\}' =  2 \frac{\omega r_v}{c}  
   (\cos\theta_0)'.
\end{equation}
Here the prime denotes $d/dx$, and small terms like $\omega\alpha,\, 
\omega\beta$ were omitted. The formal solution of this equation can be 
written as
\begin{eqnarray}
\label{betax}
   \beta(x) = \beta(0) & - & \int_{0}^{x} \alpha'(x)
   (x\theta'_0/\sin\theta_0)^2 dx \nonumber \\
   & - & (\om r_v/c) \int_{0}^{x} [x/\cos^2 
   (\theta_0/2)] dx.
\end{eqnarray}
It is easy to see that the last term is divergent for $x \rightarrow 0$, while 
the integral with $\alpha'(x)$ is convergent due to the 
presence of $\theta'(x) \sim$ exp $(-x)$. Thus, far from the vortex core we
simply have 
\begin{equation}
\label{beta}
    \beta (x) \simeq -(\om r_v/c) x^2 = - k r_v x^2,
\end{equation}
valid in the region $1 \ll x \ll (1/kr_v)^{1/2}$.
$g(r)$ then reads
\begin{equation}
\label{g(r)}
    g(r) \simeq \frac{r_v}{r} - k r_v \left(\frac{r}{r_v}\right) .
\end{equation}
On the other hand, in the region $r_v \ll r \ll (r_v/k)^{1/2}$ the 
arguments of the Bessel functions in (\ref{eq20}) are small: $kr \ll 
(kr_v)^{1/2} \ll 1$. Using the asymptotics $J_1(kr) = kr/2, Y_1(kr) = - 
2/\pi kr$, one can rewrite Eq. (\ref{g(r)}) in the form 
\begin{eqnarray}
\label{grleft}
   g(r) & \simeq & \mbox{const} 
   \left\{\frac{kr}{2} + \frac{\pi kr_v}{4} \left(- 
       \frac{2}{\pi kr}\right)\right\} \nonumber \\
   & \sim & J_1(kr) + \frac{\pi kr_v}{4} Y_1 (kr),
\end{eqnarray}
which gives us the long-wave approximation of the scattering amplitude: 
$\si_{\mp1} \rightarrow \pm\pi k r_v/4$. 
This is in good agreement with the numerical data
as we will see in the next section.

\section{Numerical diagonalization and scattering data calculation}
\label{sec:results}
%


A numerical diagonalization of the same model we are considering here 
was also already performed by Wysin and V\"olkel. \cite{Wysin95}
These authors found, for example, the splitting of the $m\neq0$ modes for OP 
vortices, as discussed above, and the lack of such a splitting for IP vortices.
For a detailed description of the eigenvalues and eigenfunctions of
(\ref{model}) we therefore refer to their paper.
For the rest of this section we relate some points which were not 
discussed in  Ref.\ \onlinecite{Wysin95} including
a short description of discreteness effects. 
Our main interest however, lies in
the scattering of magnons due to the presence of a vortex. 

We consider a circular system of radius $L$ 
on a square lattice with either fixed
(Dirichlet) or free (Neumann) boundary conditions. Static vortex
solutions were obtained using a relaxation procedure similar to that
described in Ref.\ \onlinecite{Wysin95}, or, in slightly more detail, in 
Ref.\ \onlinecite{Schnitzer_thesis}.

One problem with the diagonalization of 2D systems is that the 
number of sites in the system and therefore the dimension of
the matrix to diagonalize grows very rapidly with the linear size of
the system.
Already for a rather small system with radius $L=20a_0$  this matrix has 
dimension $2528\times2528$. If we use a conventional numerical method which 
needs to store the whole matrix in the memory of the computer, this is 
already more or less the size limit for matrices which can practically be 
diagonalized. In order to go further, we applied a special method 
which takes advantage of the structure and sparseness
of the matrix. Details are described in Appendix A.
Using this method we calculated the lowest eigenvalues and 
eigenvectors for systems up to $L/a_0=100$ 
corresponding to a matrix with dimensions $62856\times62856$.

\begin{figure}
\psfig{file=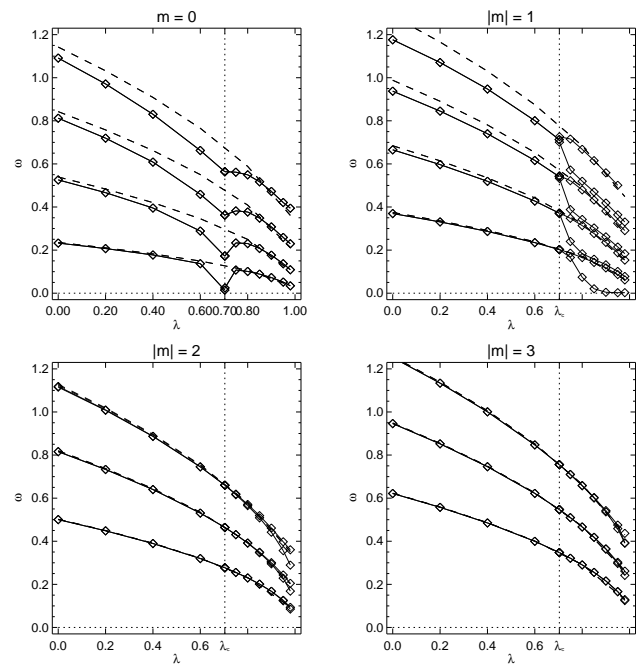,width=3.5in,angle=0.0}
\caption{Eigenvalue spectrum as a function of the anisotropy parameter 
   $\lambda$ and the ``quantum number of angular momentum'' $m$.
   Calculated for a circular system of radius $L=20$ with fixed 
   boundary conditions. The dashed lines are eigenvalues with no
   vortex in the system.}
\label{fig_ew}
\end{figure}

The eigenvalue spectrum we obtained for fixed boundary conditions 
is plotted in Fig.\ \ref{fig_ew}, as a function of the anisotropy parameter 
$\la$, for different values of the angular momentum quantum number $|m|$. 
For comparison, we have also calculated
the eigenvalues for a system linearized around the ferromagnetic
ground state of the model (i.e.\ {\em without} a vortex being present
in the system). These data are the dashed lines in Fig.\ \ref{fig_ew}.
$\la_c\approx0.70$ is the critical value of anisotropy which separates
the different regimes of stability of IP- and OP-vortices.
As can easily be seen, the mode which is responsible for this
transition is the $m=0$ mode (the lowest branch in Fig.\ \ref{fig_ew}), 
which becomes soft at $\la_c$.
Mainly the $m=0$ and $|m|=1$ modes are sensitive to the IP-OP-transition. 
The eigenvalues above $\la_c$ are split where for fixed boundary 
conditions the lower branch corresponds to negative $m$ and the upper 
branch to positive $m$.

For $\lambda<\lambda_c$, only eigenmodes with {\em odd} $|m|$ are twofold 
degenerate, as predicted in the continuum theory. 
For {\em even} $|m|$ the expected degeneracy is slightly removed. 
This is a discreteness effect, due to the fourfold symmetry
of the underlying square lattice:
eigenvectors with even $|m|$ tend on the one hand to align with a coordinate 
system parallel to the symmetry axes of the underlying lattice,
and on the other hand with a coordinate system parallel to the lattice 
diagonals. 
The latter eigenmodes oscillate with a slightly higher frequency.
 
\begin{figure}
\psfig{file=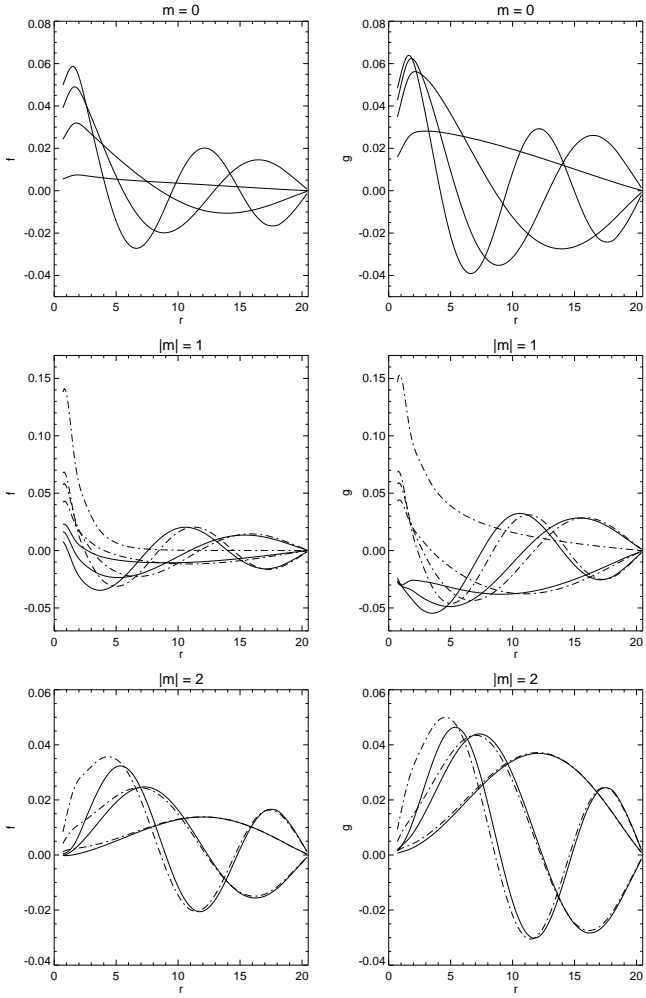,width=3.5in,angle=0.0}
\caption{Radial part of eigenfunctions for the lowest eigenvalues 
   at $\la=0.9$ using fixed boundary conditions.
   Solid and dashed lines correspond to eigenfunctions with positive $m$ and 
   negative $m$, respectively.}
\label{fig_ev}
\end{figure}

The classification of the eigenvalues according to values of $|m|$
was done by expanding the eigenvectors in a Fourier series similar
to (\ref{eq13}, \ref{eq14}) (excluding the time dependent part).
For $\lambda=0.9$ some of the radial profiles of eigenvectors 
obtained this way are plotted in Fig.\ \ref{fig_ev}.
The numerical computation always gives a complex eigenvector.
For odd $|m|$, the real and imaginary radial profiles coincide
with a high numerical accuracy.
For even $|m|$ however, as a consequence of the discreteness  effect
already described above, the radial eigenfunctions of the real and 
imaginary part differ slightly, especially near the center of the vortex.
(Fig.\ \ref{fig_ev} only shows a mean value in these cases.)
Also, for even $|m|$, the sense of rotation of the eigenmode can only
approximately be determined:  modes in the continuum theory which
belong to positive and negative $m$ intermix in the discrete system
for even $|m|$. 
This second effect is more pronounced for larger $|m|$ and larger
system sizes.
As a consequence, these eigenmodes are not in the separated 
form of Eqs.~(\ref{eq13}) and (\ref{eq14}).

The main point in our approach is to calculate the scattering matrix
$S_m(k)$ and to use its values for the investigation of general
properties of the magnon modes in the presence of the vortex. For the
extraction of the $S_m(k)$ dependence on the basis of Eq. (\ref{gen_bc},
\ref{eq20}) we only need values of eigenfrequencies for one boundary
condition for different values of the system size. 
%
Concretely, we used the fixed
b.c.\ ($b=0$ in (\ref{gen_bc}))  leading to the simple formula
\begin{equation}
\label{si_m}
   \si_m(k_i) = - \frac{J_m(k_i L)}{Y_m(k_i L)} 
\end{equation}
with $k_i=k_i(\om_i)$ according to the dispersion law (\ref{eq5}).

We calculated the scattering data for four different values of
$\la$, namely $\la=0.80$, $0.85$, $0.90$ and $0.98$, and for
different system sizes up to $L/a_0=100$. These data are presented in
Fig.\ \ref{fig_sd} as a function of $\ka=kr_v$.
We note that the points $\si_m(\ka)$ fit well to one curve for 
different $L$ and $\la$, at least for small values of 
$ka_0$. (Strictly speaking, the inequality $ka_0\ll1$ should be satisfied,
but even for $ka_0 \approx 0.4$ the
discrepancies among different values of $\la$ are negligible for the 
mainly interesting case $|m| = 1$.)


For the general boundary conditions (\ref{gen_bc}) the frequency spectrum
is determined by the dimensionless equation
\begin{eqnarray}
\label{aJ_m}
   & a &[J_m(kL) + \si_m(kr_v) Y_m(kL)]  \nonumber \\
    {}+  kr_v & b &[J_m'(kL) + \si_m(kr_v) Y_m'(kL)] = 0.
\end{eqnarray}
We start our discussion for the modes with $|m|>1$ and $m=0$.
The translational modes ($|m|=1$) will be considered in the next section.

For $|m|>1$ and fixed b.c.\ ($b=0$ in (\ref{gen_bc})), taking into account
$\si_m<0$, 
it can easily be seen that Eq.\ (\ref{aJ_m}) has no solution in the region
$kL \ll 1$ where $J_m(z)\sim (z/2)^m$ and $Y_m(z)\sim -(2/z)^m/\pi$. 
The $n$'th root can be written as
$kL=z_{n,m}$ with $y_{m,n}<z_{m,n}<j_{m,n}$. Here $y_{m,n}$ and
$j_{m,n}$ are the $n$'th zeros of $Y_m(z)$ and $J_m(z)$, respectively.
Because $j_{m,n}$ and $y_{m,n}$ are smaller than $\pi(n+m)/2$
for large $n$ or $m$, the low-lying modes have the values
$k_{n,m}\simeq \max(n,m)/L$. This means $\si_m \ll 1$ for 
$r_v/L  \ll 1$, hence the eigenfrequencies can approximately
be described by the zeros of $J_m(z)$ (which have approximately the same 
values as for the system without vortex). This can also be observed very 
clearly  in Fig.\ \ref{fig_ew}.

For the case $a\sim b$ in the b.c.\ (\ref{gen_bc}) the same holds because of
the presence of the small parameter $kr_v$ in the term with the
derivatives. For $a=0$ (Neumann b.c.), however, the situation changes.
First of all, the formula for $k_{n,m}$ can be presented in the
form $k_{n,m}=j'_{m,n}/L$ where $j'_{m,n}$ is the
$n$'th zero of $J_m'(z)$. More importantly, a new root of (\ref{aJ_m}) 
appears, with $k=k_{0,m} \ll 1/L$ corresponding to the lowest mode with 
$|m|>1$.
In order to show this we use again the asymptotics of $J_m(z)$
and $Y_m(z)$  for $z \ll 1$. 
Applying the formula $\si_m=-A_m (kr_v)^{\alpha}$, one easily obtains
\begin{equation}
\label{r_vk}
   r_vk_{0,m} = \left(\frac{A_m}{\pi}\right)^{p/2m} 
\left(\frac{2r_v}{L}\right)^p \quad,\quad
   p = \frac{1}{1-\al/2m} > 1.
\end{equation}
The corresponding frequency $\om_{0,m}\approx ck_{0,m}$
decreases faster than $1/L$ for $L\to\infty$.

The eigenmodes with $m=0$  can be described in the same way
as for $|m|>1$. Only the values $k_{n,0}=(1/L) j_{0,n}$ or
$k_{n,0}=(1/L) j_{0,n}'$ are present in this case, and 
$k_{n,0}\sim 1/L$ for general boundary conditions. 
The only exception is the lowest
mode for  Neumann b.c.  For this mode the analysis of the asymptotics
of cylindrical functions gives exactly $\om=0$. This value corresponds
to the Goldstone mode due to the presence of a rotational symmetry
in spin space which is still exact for the finite system.

\section{Translational modes and collective variable approach}
This section deals with the analysis of the translational modes $|m|=1$
and the application of their properties for the construction of equations 
of motion for the vortex dynamics.
 
\begin{figure}
\psfig{file=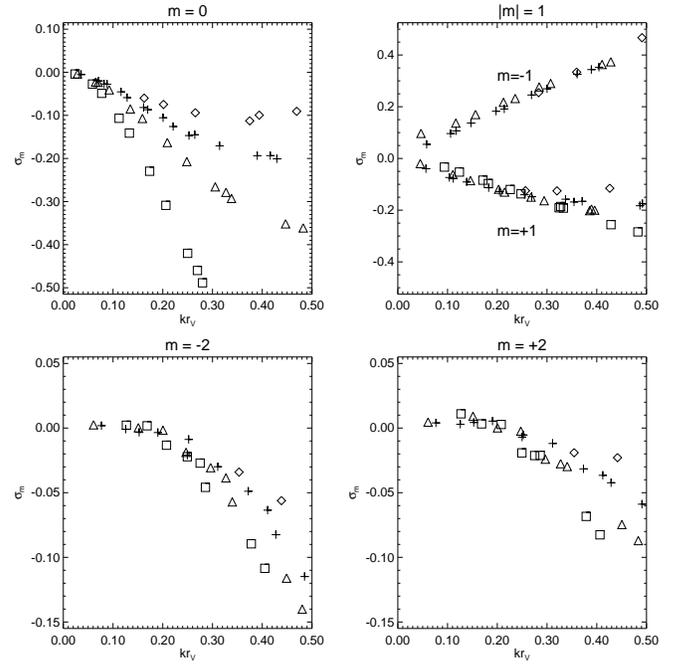,width=3.5in,angle=0.0}
\caption{Scattering data for different values of $\lambda$
   ($\Box=0.80$, $\triangle=0.85$,$+=0.90$, $\Diamond=0.98$) and
   different system sizes in the range $L=15\ldots100$.}
\label{fig_sd}
\end{figure}

From Fig.\ \ref{fig_sd} one can see that the scattering parameter for $|m|=1$ 
have a linear dependence $\sigma_{\pm1}(k) = \mp A_1 kr_v$. 
The long-wave approximation (\ref{grleft}) gives 
$A_1 = (\pi/4)$, which is in good agreement 
with Fig.\ \ref{fig_sd}. Using the asymptotics $J_1 \simeq z/2$, $Y_1 \simeq - 2/\pi z$ 
for $z \ll1$, we get for the frequency of the lowest translational
mode $\om_0 = \pm cr_v/L^2$ where the positive sign corresponds to
Neumann and the negative sign to Dirichlet boundary conditions, 
respectively. Using $c r_v = a^2_0JS \sqrt{\lambda}$ and replacing 
$\sqrt{\lambda} \rightarrow 1$, which is natural in the continuous small 
anisotropy approximation, we obtain $\om_0=\pm JSa_0^2/L^2$.

This analytical result is in agreement with the eigenvalue data
obtained numerically.
However, in the discrete system, $\om_0$ does actually not tend to zero 
for $L\to\infty$ but to a finite constant $\om_p$.  
Using the collective variable approach this constant can easily be
understood as a pinning effect as we would like to show now.
It is known that the dynamics of the OP vortex can  be described in lowest
approximation by the so-called Thiele Equation \cite{Thiele73,Huber82}
\begin{equation}
\label{vecG}
   -\vec G \times \dot{\vec X} = \vec F.
\end{equation}
Here $\vec G$ is the gyrovector 
$\vec{G}=(2 \pi q p S/a_{0}^2) ~ \hat{z}$, where $\hat{z}$ is the unit
vector along the hard ($z$) axis,
$\vec X$ the position of the vortex center,
and $\vec F$ the force acting on the vortex.
In the case where the vortex is only slightly displaced from the center of 
the system, Thiele's Eq.\ just describes the dynamics of the lowest 
translational mode. There are two types of forces acting on the vortex. 
The first one is due to an image vortex which resides at the position
($L^2/ X^2)\vec X$, $X=|\vec X|$, and the second one is due to the pinning 
potential of the underlying lattice. Putting both together we obtain
\begin{equation}
\label{vecGtime} 
  -\vec G\times \dot{\vec X} = -\frac{2\pi JS^2qq_i\vec X}{L^2- X^2} 
                                - k_p \vec X,
\end{equation}
where $q_i$ is the vorticity of the image vortex.
We have assumed that the pinning potential is harmonic in lowest
approximation leading to a linear force with spring constant $k_p$.
In the limit $X\to0$, Eq.\ (\ref{vecGtime}) has the 
solution $X_1(t) = X\cos \om_0 t$,  $X_2(t) = X \sin \om_0 t$ with
\begin{equation}
\label{vd2:MS:w2}
   \om_0 = -\frac{q q_iJSa_0^2}{L^2} - \frac{k_pa_0^2}{2\pi S}
\end{equation}
which, apart from the constant part, agrees with the dependency we have found
above. 
In order to calculate $k_p$ from the numerical eigenvalue data 
one can use the fact that Dirichlet
boundary conditions generate an image of the same vorticity
while Neumann boundary conditions generate an image  of 
opposite vorticity. Therefore from (\ref{vd2:MS:w2}) two equations 
follow:
\begin{eqnarray}
\label{vd2:MS:wGSM}
   \om_{0,Dirichlet} & = & -\frac{JSa_0^2}{L^2} - \frac{k_pa_0^2}{2\pi S}
 \nonumber \\
   \om_{0,Neumann} & = & +\frac{JSa_0^2}{L^2} - \frac{k_pa_0^2}{2\pi S}. 
\end{eqnarray} 
Adding them leads to the following formula for $\om_p$ which is
independent of the concrete size of the system:
\begin{equation}
\label{Neum}
   \om_p \,\equiv\, -\frac{k_pa_0^2}{2\pi S} 
   \,=\, \frac{\om_{0,Dirichlet} + \om_{0,Neumann}}{2}.
\end{equation}
Assuming a pinning potential of the form 
$E_p[\cos(2\pi x/a_0)+\cos(2\pi y/a_0)+2]/2$, $\om_p$ is connected to
the pinning energy $E_p$ by the formula
\begin{equation}
\label{E_p}
    E_p = \frac{S|\om_p|}{\pi} .
\end{equation}
We have evaluated (\ref{Neum}) and (\ref{E_p}) for some values of $\la$, 
cf.\ Table \ref{table1}. 
Naturally, the pinning energy decreases with decreasing anisotropy
because the size of the OP-structure increases strongly, making
discreteness effects less and less important.

The pinning energy can also be measured directly,
comparing the energy of static vortices centered on different lattice
coordinates. These numbers, labeled $E_p^{direct}$ in
Table \ref{table1}, compare very well with $E_p$ due to formula (\ref{E_p}).

Next, we discuss the higher modes obeying the $|m|=1$ symmetry. 
Due to the general features 
discussed in the previous section, these modes can be considered as 
doublets, with frequencies $\om_n > 0$ and $\om_{n+1} < 0$ with 
$|\om_{n+1}|\simeq \om_n$. For general b.c.\ the mean frequency can be 
written as
\begin{equation}
   \label{overl}
   \overline{\om}_n \simeq cx_n/L,
\end{equation}
where $x_n$ is the $n$-th root of the equation $aJ_1(x) + b x(r_v/L) J'_1 
(x) = 0$. For fixed and free b.c.\ one has $\overline{\om}_n = cj_{1,n}/L$ 
and $\overline{\om}_n = cj'_{1,n}/L$, respectively. 

We concentrate on the lowest doublet with the frequencies $\om_1$ and 
$\om_2 \simeq - \om_1$ on the following reasons: (i) these modes can be 
compared with long-time computer simulations of the vortex motion and (ii) 
they can be used for the calculation of parameters in equations of 
the vortex motion. In the lowest approximation in $r_v/L$ one can write 
\begin{equation}
   \label{om/SJ}
   \frac{\overline{\om}_1}{SJ}  \simeq 
   2x_1\sqrt{1-\lambda} \frac{a_0}{L},
\end{equation}
where $x_1 = j_{1,1} \simeq 3.832$ for fixed b.c.\ and $x_1 \simeq 
j'_{1,1} \simeq 1.8412$ for free ones. For general b.c.\ $x_1$ lies 
between these values. Note, however, that for $a \approx b$ the value of 
$x_1$ is still close to $j_{1,1}$, and the ``switching'' to the value 
$j'_{1,1}$ occurs only for small $a \le b (a_0/L)$. We remind that 
$\overline{\om}_1$ does not depend on the scattering data in the first 
approximation in $a_0/L$.
The use of scattering data becomes important however for the calculation of the 
doublet width $\Delta\om = |\om_2| - \om_1$. Considering the leading 
approximation $kr_v\sim a_0/L$, the width can be presented in the 
form
\begin{equation}
   \label{delta}
   \frac{\Delta\om}{SJ} = \frac{\pi\sqrt{\la}}{2} (\frac{a_0}{L})^2
   \frac{x_1^2 (ax_1 - b) Y_1 + x_1bY_0}{x_1(x_1a-b)J_0+[(2-x_1^2) 
   b-x_1a]J_1},
\end{equation}
where $J_0, J_1, Y_0$ and $Y_1$ all have the argument $x_1$. One can see 
that $\Delta\om$ is practically independent of the anisotropy (we will omit 
the coefficient $\sqrt{\la}$ below), and it is inversely proportional to 
$L^2$. This is the same dependence as for the lowest translational mode
$\om_{0} \simeq JS (a_0/L)^2$, but the coefficients are different. 

\begin{figure}
\psfig{file=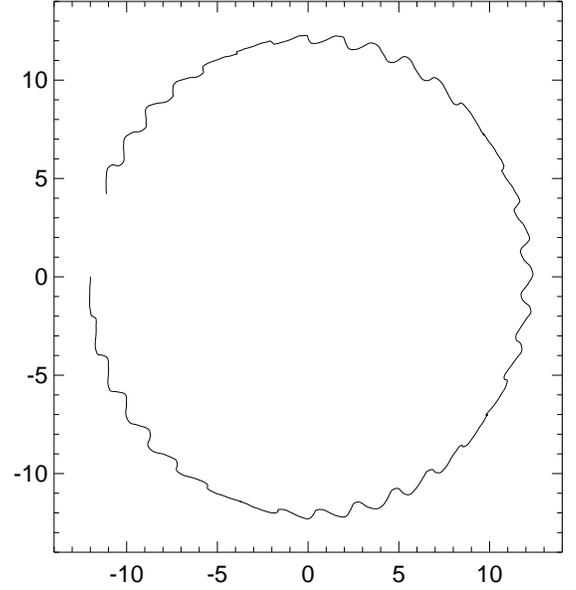,width=3.5in,angle=0.0}
\caption{Trajectory of a vortex obtained by numerically 
    integrating the Landau Lifshitz equation in time.
    The simulation was performed on a circular system with radius $L=36a_0$
    and free boundary conditions.}
\label{fig_traj}
\end{figure}

Next we compare our results for $\om_{0}$ and the lowest doublet $\om_{1,2}$ 
with data from recent computer simulations \cite{Mertens97} for the motion 
of one OP-vortex on large circular systems (up to $L  = 72 a_0$) with free 
boundaries. The Landau-Lifshitz Eq. was integrated for a square lattice with 
$a_0 = 1$ and the Hamiltonian (\ref{model}), where $J = S = 1$ and $\lambda 
= 0.9$. The vortex center performs oscillations around a mean trajectory 
$\vec{X}^0(t)$ which is a circle with radius $R_0$  around the circle center 
(Fig.\ \ref{fig_traj}). The mean trajectory can be interpreted as a stationary solution of 
the Thiele Eq. (\ref{vecG}), where the vortex is driven by the interaction 
with its image vortex. The rotational frequency $\om_0$ goes to a constant 
for $R_0 \rightarrow 0$; the extrapolated value $0.201\;10^{-3}$ agrees 
well with our result $0.193 \; 10^{-3}$ (both for $L = 72$).

The Fourier spectrum of the oscillations around $\vec{X}^0(t)$ shows a 
doublet $\om'_{1,2}$ with about equal amplitudes, and phases $\mp   
\pi/2$ between the two components. As the spectra in Ref.\ \onlinecite{Mertens97} 
were evaluated in a moving polar coordinate frame we must add and subtract 
$\om_0$ in order to compare with our results in Eqs.\ (\ref{om/SJ}) and 
(\ref{delta})
 \begin{equation}
   \label{om_12}
   \om_{1,2} = \om'_{1,2} \pm \om_0.
\end{equation}
The data agree very well, e.g. for $L = 72$ and $R_0 = 16.11$ the 
frequencies $\om_0 = 2.05\;10^{-4},\, \om'_1 = 1.546\;10^{-2}$, 
and $\om'_2 = 1.661 \; 10^{-2}$ were observed which yields 
$\overline{\om} = 1.604 \; 10^{-2}$ for the mean and $\Delta\om = 0.74 
\; 10^{-3}$ for the difference. This agrees with our 
theoretical values $\overline{\om} = 1.617 \; 10^{-2}$ and $\Delta\om = 
0.7657 \; 10^{-3}$ within 0.8 and 4\%, respectively. 
 
Note that the very good agreement of our normal mode approach with the data 
from the simulations is not a trivial result: we have calculated the 
frequencies appearing in the vortex dynamics for {\it free boundary 
conditions}, 
using as input the scattering data for {\it fixed boundary conditions}. 
This shows that our 
scattering theory actually works for the general case. 

We conclude that the vortex motion is accompanied by, or generates, the two 
quasi-local magnon modes with $|m| = 1$. These show up in two ways: (i) in 
the trajectory of the vortex center as oscillations around a mean 
trajectory, (ii) as oscillations of the dynamic parts of the vortex 
structure (see below).

There remains the question whether the trajectory of the vortex center can 
be obtained from an equation of motion. 
If the rigid-shape assumption of Thiele is dropped by 
allowing for a deformation due to the velocity, \cite{fn2}
a $2^{nd}$-order equation of 
motion can be derived \cite{Wysin94a} which exhibits an additional term 
$M\ddot{\vec{X}}$ with an effective vortex mass $M$. The additional term 
causes cycloidal oscillations \cite{fn3} 
with frequency $\om_c = G/M$ around a mean trajectory. For 
the vortex mass $M \sim \ln L$ was predicted, which was not confirmed by 
computer simulations, though Refs.\ \onlinecite{Mertens94,Voelkel94}. Recently Wysin 
\cite{Wysin96} proposed to calculate $M$ by using the two lowest 
eigenfrequencies, $\om_0$ and $\om_1$ in our notation. His formula $M_W = 
G/(\om_0 + \om_1)$ gave a linear $L$-dependence when the frequencies from 
his numerical diagonalization were inserted. The same dependence is 
obtained (in $O(a_0/L))$ if we insert our analytical results for the 
frequencies.

However, meanwhile new simulations had been performed which resulted in the 
observation of the above mentioned doublet $\om_{1,2}$, instead of a single 
frequency $\om_c$. This dynamics can be fully described by a 
$3^{rd}$-order equation of motion \cite{Mertens97} which was derived by a 
collective variable theory, starting from a generalized traveling wave 
ansatz
\begin{equation}
   \label{vec_sr}
   \vec{S}(\vec{r},t) = 
   \vec{S}(\vec{r} - \vec{X}, \dot{\vec{X}}, \ddot{\vec{X}}),
\end{equation} 
where the vortex shape 
is assumed to depend also on the acceleration. In fact, this dependence can 
be seen in simulations with free b.c.\ when one considers the spin 
configurations at the turning points of the trajectory where the 
acceleration is maximum while the velocity is small. \cite{Mertens97}

The $3^{rd}$-order equation has the form 
\begin{equation}
   \label{G_3times} 
   \vec{G}_3 \times \dddot{\vec{X}} + M\ddot{\vec{X}} - \vec{G}\times 
   \dot{\vec{X}} = \vec{F}
\end{equation}
with $\vec{G}_3 = G_3\vec{e}_z$ and a new parameter $G_3$. We note that this 
is the most general $3^{rd}$-order equation for the given easy-plane 
symmetry. In Ref. \onlinecite{Mertens97}, $G_3$ was defined as an integral over 
$\vec{r}$ which could not be performed because the dynamic vortex structure 
is not known analytically for the vortex core. But the size dependence $G_3 
\sim L^2$ was obtained from the outer region of the integral. We will see 
below that our theory allows for the calculation of $G_3$ and $M$.

We consider Eq. (\ref{G_3times}) for the case of vortex motion close to the 
center of the circle where the image force is approximately linear
\begin{equation}
   \label{F-2}
   \vec{F} = - 2 \pi qq_i \vec{X}/L^2.
\end{equation}
Then Eq. (\ref{G_3times}) can be solved by a harmonic ansatz, where the 
frequency $\om$ fulfills
\begin{equation}
   \label{G_3om}
   -G_3\om^3 - M \om^2 + G\om = - 2 \pi qq_i/L^2.
\end{equation}
The parameters can be obtained from the three roots $\om_i$, using Vieta's 
rules
\begin{eqnarray}
   \label{om_0+}
   \om_0 + \om_1 + \om_2 &=& - M/G_3\\
   \label{om_0/}
   \om_0\om_1 + \om_0\om_2 + \om_1\om_2 &=& - G/G_3\\
   \label{om_0/1}
   \om_0\om_1\om_2 &=& 2 \pi qq_i/(G_3L^2).
\end{eqnarray}
We now identify $\om_0$ with the Goldstone mode $-qq_i/L^2$, neglecting the 
frequency $\om_p$ from the pinning force since Eq. (\ref{G_3times}) was derived 
in the continuum limit; $\om_1$ and $\om_2 < 0$ are identified with the 
lowest doublet calculated in Eqs.\ (\ref{om/SJ}, \ref{delta}).

From Eq. (\ref{om_0/1}) we obtain \cite{fn4}
\begin{equation}
   \label{G_3frac}
   G_3 = \frac{2\pi}{\overline{\om}^2}
\end{equation}
with $\overline{\om} = \sqrt{\om_1 |\om_2|} \simeq \frac{1}{2} (\om_1 
+ |\om_2|)$, which gives
\begin{equation}
   \label{G_3fracpi}
   G_3 = \frac{\pi}{2x_1^2 (1 - \la)} L^2 .
\end{equation}
The result $4.634 L^2$ for free b.c.\ agrees very well with $4.67 
L^{2.002}$ for large $L$ obtained from the simulation data. \cite{Mertens97}  
For fixed b.c., $G_3 = 1.07 L^2$ is about four times smaller.

From Eq. (\ref{om_0+}) we obtain
\begin{equation}
   \label{M/frac}
   M = \frac{2\pi}{\overline{\om}^2}\, \left(\Delta\om + \frac{qq_i}{L^2}\right)
\end{equation}
with $\Delta\om = |\om_2| - \om_1$. Here all $L$-dependencies just cancel in
the lowest approximation on $L$, 
thus the vortex mass in Eq. (\ref{G_3times}) is {\it independent} of the 
system size, as already obtained from the simulations. \cite{Mertens97}
Inserting Eq. (\ref{delta}) we get for free b.c.
\begin{equation}
   \label{M/fracpi}
   M = \frac{\pi}{2 (1 - \la)} \left(\frac{\pi}{2} \frac{x_1Y_0 - Y_1}{(x_1^2 - 
   1)J_1} - \frac{1}{x_1^2}\right).
\end{equation}
The numerical value $M = 14.74$ agrees well with $M = 15$ which we have 
extrapolated for $R_0 \rightarrow 0$ from the data in Ref. 
\onlinecite{Mertens97} for three different system sizes $(L = 24, 36, 72)$. We 
note that the vortex mass (\ref{M/fracpi}) is in the same order of magnitude
 as 
the two-dimensional soliton mass $M = E_0/c^2$ in Ref. \onlinecite{Ivanov89}, where $E_0 = 4 \pi$ 
(in units $JS^2$) is the Belavin-Polyakov energy.

For fixed b.c. 
\begin{equation}
   \label{M/fracpi2}
   M = \frac{\pi}{2 (1 - \la)} \left(\frac{\pi}{2} \frac{Y_1}{x_1J_0} + 
   \frac{1}{x_1^2}\right)
\end{equation}
yields $M = 7.661$, which is about one half of the above value for free 
boundary condition.

Thus in the equation of motion (\ref{G_3times}) only the gyrovector $\vec{G}$ is an
intrinsic property of the vortex. The quantities $M$ and $G_3$, which are
connected to the quasi-local modes with frequencies $\omega_{1,2}$, are
determined by the whole system ``vortex plus magnons'' which includes
the geometry of the system and the boundary conditions. For $G_3$ this
is obvious because it strongly depends on $L$; but $M$ does not depend
on $L$ (in the lowest order), however, it depends on the boundary
conditions. On the other hand, the S-matrix naturally is determined 
only by the region of the vortex core and can thus be used for the
calculation of the parameters $M$ and $G_3$ for arbitrary geometry 
and boundary conditions.

Finally we point out that Eq. (\ref{G_3times})
belongs to a whole hierarchy of equations of motion which can be derived 
\cite{Mertens97} by taking into account higher and higher time derivatives 
of $\vec{X}(t)$ in the generalized traveling wave ansatz (\ref{vec_sr}). 
Moreover,  only the odd-order equations of this hierarchy 
represent valid approximations, because the even-order equations have a very 
weak leading term,  e.g. $M\ddot{\vec{X}}$ in the $2^{nd}$-order equation. 
Therefore the solutions of the latter equations are qualitatively different 
from the solutions of the odd-order equation and are in fact not confirmed 
by the simulations (cf. the above discussion of the single frequency $\om_c$ 
and the doublet $\om_{1,2}$, which exhibit quite different dependencies on 
the system size). The $5^{th}$-order equation of the hierarchy predicts a 
second doublet $\om_{3,4}$ which was in fact observed in the simulations, 
but only for specially designed initial and boundary conditions because the 
amplitudes of this doublet are very small. \cite{Mertens97} Thus the 
$3^{rd}$-order equation represents already a very good approximation. The 
observed additional doublet naturally also appears in the results of our 
numerical diagonalization in Section \ref{sec:results} and its 
frequencies can be calculated 
by $\overline{\om}_n$ and $(\Delta\om)_n$, see Eq. (\ref{overl}) and below.

\section{Conclusion}
\label{sec:conclusion}

We have developed a general theory which allows calculation of the magnon 
modes of a circular easy-plane ferromagnet in the presence of an 
out-of-plane vortex. The method consists of a combination of numerical 
diagonalization of the discrete system with analytical calculations in the 
continuum limit. The frequencies of the magnon modes can be expressed in 
terms of the functions $\sigma_m(kr_v)$, which are independent of the 
magnetic coupling constants, the system size and the boundary conditions. 
The $\sigma_m$ describe the intensity of the magnon scattering due to the 
presence of the vortex.

The translational modes with $|m| = 1$ are particularly interesting for two 
reasons: (i) Their frequencies are identified in the vortex motion which was 
observed in simulations where the Landau-Lifshitz Equation was integrated 
for the  circular discrete spin system. (ii) Using these frequencies, one 
can calculate the two parameters of a $3^{rd}$-order equation for the vortex 
motion (a generalization of the Thiele Eq.), which was derived by a 
collective variable theory starting from a generalized traveling wave 
ansatz. Our calculated parameters agree very well with those obtained by 
describing the simulations using the $3^{rd}$-order equation of motion. Both 
parameters, the vortex mass $M$ and the $3^{rd}$-order gyrocoupling constant 
$G_3$, depend strongly on the boundary conditions. This is due to the fact 
that both the in-plane and the out-of-plane structure of the moving vortex 
are not localized. In fact, the dynamic parts of the vortex structure 
oscillate with the frequencies of the translational modes with $|m| = 1$.

\acknowledgments
BAI thanks the University of Bayreuth for kind hospitality
where part of this work was performed. GMW acknowledges the
support of NSF Grant No.\ DMR-9412300. We acknowledge
Ms.\ Sigrid Glas for help in preparation of the manuscript.

\begin{appendix}

\section{Numerical diagonalization of Hamiltonian matrices}

If a classical Hamiltonian system is been linearized around a static solution,
an eigenvalue problem  
\begin{equation}
\label{app2:1}
    {\bf A} \vec z = \lambda \vec z
\end{equation}
results with a coefficient matrix $\bf A$ fulfilling the generalized 
symmetry relation
\begin{equation}
\label{app2:2}
   \bf A^T J = J^T A.
\end{equation}
$\bf J$ is the ``symplectic unit matrix'', i.e.
\begin{equation}
\label{app2:3}
   {\bf J} = \left( \begin{array}{cc} {\bf 0} & {\bf 1} \\ -{\bf 1} 
                   & {\bf 0} \end{array}\right)  .
\end{equation}
For obvious reasons the matrix $\bf A$ is called a ``Hamiltonian matrix''.
It is easy to see that $\bf A$ can always be written as $\bf JH$ 
where $\bf H$ is a symmetric matrix.
Note that the eigenvalues and eigenvectors of $\bf A$ are complex in 
general.

For the most general case, Hamiltonian matrices have no special properties
which could profitably be used in the numerical diagonalization.
An exception, however,  is the case of a positive definite matrix $\bf 
H$, for which (\ref{app2:1}) is equivalent to an Hermitian eigenvalue problem
with pure imaginary eigenvalues.
In order to show this, we note that a positive definite matrix can always
be written as
\begin{equation}
\label{app2:4}
   \bf H = L L^T,
\end{equation}
the so-called Cholesky decomposition of $\bf H$, 
cf.\ Ref.\ \onlinecite{StoerBulirsch}.
Substituting this in (\ref{app2:1}) and additionally defining
$\lambda:=i\omega$ we obtain
\begin{equation}
\label{app2:5}
   (i{\bf L^TJL})({\bf L^T}\vec z) = -\omega ({\bf L^T}\vec z).
\end{equation}
Now, $\bf J$ is antisymmetric and therefore  $i\bf L^TJL$ is Hermitian.

Unfortunately, it is somehow difficult to use this equivalence directly
for the purpose of numerical diagonalization.
The Cholesky decomposition of a sparse symmetric matrix can be computed 
easily only for certain cases, otherwise it is very time 
consuming. For this 
reason we did not directly use formula (\ref{app2:4}). Instead we 
utilized a method called Wielandt's version of the inverse iteration 
procedure. 
The basic strategy is to multiply a (randomly chosen) vector over and over 
by the inverse of the spectral-shifted matrix $\bf A$, i.e.\ 
$({\bf A}-\bar\lambda {\bf I})^{-1}$. The resulting series of vectors 
converges to an eigenvector of $\bf A$, usually the one corresponding to 
the eigenvalue closest to the chosen spectral shift $\bar\lambda$. 
The details of the method can be found in Ref.\ \cite{StoerBulirsch}.
For Hamiltonian matrices, supposing a positive definite matrix $\bf H$
and therefore imaginary eigenvalues and an imaginary spectral shift
$\bar\lambda = i\bar\omega$,
Wielandt's inverse iteration amounts to the following iteration formula:
\begin{equation}
\label{app2:6}
%
%
%
   \left(\begin{array}{cc} {\bf H}  & -\bar\omega {\bf J} \\ 
                           \bar\omega {\bf J} & {\bf H} 
         \end{array}\right)
   \left(\begin{array}{c} \vec{x}^{(j+1)} \\ \vec {y}^{(j+1)} 
         \end{array}\right) =
  - \left(\begin{array}{c} {\bf J} {\cal P} \vec{x}^{(j)} \\ 
             {\bf J} {\cal P} \vec{y}^{(j)} \end{array}\right) .
\end{equation}
$\vec{x}^{(j)}$ and $\vec{y}^{(j)}$ are the real and imaginary part of the 
eigenvector in the $j$'th iteration step. The initial vector 
$(\vec{x}^{(0)},\vec{y}^{(0)})$ must be choosen randomly. 
The matrix ${\cal P}$ is a symplectic projection operator which is defined as
\begin{equation} 
\label{app2:8}
   {\cal P} = \sum_{k\le j} \frac{\vec y^{(k)}\vec x^{(k)}{}^T 
                 - \vec x^{(k)}\vec y^{(k)}{}^T}
                   {\vec x^{(k)}{}^T{\bf J}\vec y^{(k)}}{\bf J} .
\end{equation} 
The sum runs over all eigenvectors which were computed 
in previous runs. The purpose of this operator is to avoid that the 
method converges to an already-known eigenvector.

After some iteration steps the parameter $\bar\omega$ can also be iterated, for
example according to the formula
\begin{equation}
\label{app2:7}
   \bar\omega = \frac{1}{i} \frac{\vec z^+{\bf A} \vec z}{ |\vec z|^2}
   \quad,\quad \vec z = \vec x^{(j)} + i \vec y^{(j)}.
\end{equation}
As $\vec z$ converges to an eigenvector of $\bf A$,
$\bar\lambda$ converges to the corresponding eigenvalue. 

A positive definite matrix $\bf H$ is not a principal presumption for
the inverse iteration, i.e. the above iteration formula can in principle
be generalized to include the case of non-definite matrices $\bf H$.
However, first of all, with $\bf H$, the coefficient matrix in (\ref{app2:6}) 
is also positive definite.  
This allows to use an efficient numerical method to solve (\ref{app2:6}).
To be precise we have used the NAG library function F04MBF for this purpose
which is based on a Lanzcos algorithm.
The NAG routine makes it only necessary to supply a ``matrix$\times$vector''
function. Therefore one has to store only the elements of $\bf A$ which are 
non-zero which precisely makes the method suitable for large and sparse 
matrix equation. 
A second point is that for non-definite $\bf H$ rather serious problems 
with the numerical stability of the inverse iteration arise.



\end{appendix}




\begin{table}
\label{table1}
\caption{Dependence of pinning frequency and energy on the anisotropy parameter $\la$.}
\begin{tabular}{cccc}
   $\la$ & $|\om_p|/JS$     & $E_p/(JS^2)$    & $E^{direct}_p/(JS^2)$ \\\hline
   0.80  & 0.0672           & 0.0214          & 0.0233 \\
   0.85  & 0.0181           & $5.76\;10^{-3}$ & $5.87\;10^{-3}$ \\
   0.90  & $1.88\; 10^{-3}$ & $5.98\;10^{-4}$ & $6.05\;10^{-4}$ \\
   0.98  & $8.07\; 10^{-5}$ & $2.57\;10^{-5}$ & $3.85\;10^{-6}$ \\
\end{tabular}
\end{table}

\newpage

\begin{thebibliography}{99}
\bibitem[\dagger]{fn1} For antiferromagnets this matrix was calculated by a different
   method. \cite{Ivanov96}
\bibitem[\ddagger]{fn2} A mass, due to a velocity induced shape deformation, was 
   already introduced for the solitons in 2D 
   easy-axis ferromagnets. \cite{Ivanov89}
\bibitem[\S]{fn3} These trajectories correspond to the cycloids with frequency 
   $eB/(Mc)$ of an electric charge $e$ in a plane with a perpendicular magnetic 
   field $\vec{B}$ and an in-plane electric field $\vec{E}$, where $c$ is the 
   speed of light.
\bibitem[\P]{fn4} Eq. (\ref{om_0/}) gives the same result, because a term 
   $\om_0 \Delta\om/\overline{\om}^4$ can be neglected.
\end{thebibliography}
\end{document}